\documentclass[iop]{emulateapj}
\newcommand{\kms}{km~s$^{-1}$}
\newcommand{\VLSR}{V$_{\rm LSR}$}

\shorttitle{Outflow from Orion Source I}
\shortauthors{Plambeck et al.}

\begin{document}

\title{Tracing the bipolar outflow from Orion Source I}

\author{R.~L.~Plambeck\altaffilmark{1}, M.~C.~H.~Wright\altaffilmark{1}, 
   D.~N.~Friedel\altaffilmark{2}, S.~L.~Widicus~Weaver\altaffilmark{3}, \\
   A.~D.~Bolatto\altaffilmark{4}, M.~W.~Pound\altaffilmark{4},
   D.~P.~Woody\altaffilmark{5}, J.~W.~Lamb\altaffilmark{5}, S.~L.~Scott\altaffilmark{5}}

\altaffiltext{1}{Radio Astronomy Lab, 601 Campbell Hall, University of California, Berkeley, CA 94720 USA}
\altaffiltext{2}{Department of Astronomy, 1002 W. Green St., University of Illinois, Urbana, IL 61801 USA}
\altaffiltext{3}{Department of Chemistry, Emory University, Atlanta, GA 30322 USA}
\altaffiltext{4}{Department of Astronomy, University of Maryland, College Park, MD 20742 USA}
\altaffiltext{5}{Owens Valley Radio Observatory, California Institute of Technology, P.O. Box 968,
   Big Pine, CA 93513 USA}


\begin{abstract}

  Using CARMA, we imaged the 87 GHz SiO v=0 J=2-1 line toward Orion-KL
  with $0.45\arcsec$ angular resolution.  The maps indicate that radio
  source~I drives a bipolar outflow into the surrounding molecular
  cloud along a NE--SW axis, in agreement with the model of
  \citet{Greenhill04}.  The extended high velocity outflow from
  Orion-KL appears to be a continuation of this compact outflow.  High
  velocity gas extends farthest along a NW--SE axis, suggesting that
  the outflow direction changes on time scales of a few hundred years.

\end{abstract}

\keywords{ISM: individual(Orion-KL) --- ISM: jets and outflows ---
masers ---  stars: formation } 

\section{Introduction}

Bipolar outflows are ubiquitous from young low-mass stars, but are
difficult to observe toward high-mass stars, which typically form in
crowded, physically complex regions that are quickly disrupted by
ionization fronts and stellar winds.

The nearest region of massive star formation is the Kleinmann-Low
Nebula in Orion, at a distance of about 400 pc
\citep{Sandstrom07,Menten07}.  Measurements of H$_2$O maser proper
motions led \citet{Genzel81} to suggest that two distinct outflows
originate from this region -- a low velocity (18 \kms) outflow along a
NE--SW axis, and a high velocity (30--100 \kms) flow extending roughly
NW--SE.  Both outflows were inferred to originate within a few
arcseconds of radio source~I, a young star with a luminosity of $10^4$
to $10^5$ L$_{\sun}$ \citep{Gezari98}.  The high velocity outflow also
manifests itself as lobes of shock-excited H$_2$ \citep{Beckwith78}
and as broad, weakly bipolar line wings in CO and other molecules
\citep{Zuckerman76,Kwan76,Erickson82,Chernin96}.

Source I is associated with a cluster of SiO masers.  From fits to
$2\arcsec$ resolution BIMA data, \citet{Plambeck90} found that the
J=2-1 v=1 SiO masers were clustered along two arcs offset $\sim
0.08\arcsec$ NW and SE of source~I.  \citet{Plambeck90} reproduced
this pattern with a model of maser emission from a rotating, expanding
disk, tilted at $45^{\circ}$ to the plane of the sky.  The axis of the
model disk was projected at PA $145^{\circ}$, suggesting that the high
velocity outflow emerged along its poles.

Later observations showed that the J=2-1 SiO line in the v=0
vibrational level also was masing \citep{Wright95} in a 1000 AU long
hourglass-shaped region centered on source~I.  Because the hourglass
was elongated NE--SW, it was natural to associate it with the outer
regions of the model v=1 maser disk.  SiO linewidths were $\sim
30$~\kms\ across the entire hourglass, inconsistent with Keplerian
rotation, so \citet{Wright95} suggested that the emission traced the
turbulent boundary layer between an underlying, unseen, disk and the
high velocity outflow.

More recent observations have cast doubt on this picture.  VLBA
observations of the 43 GHz v=1 SiO masers
\citep{Greenhill98,Doeleman99} showed that the brightest maser spots
were clustered in 4 groups, rather than than in a ring as predicted by
the \citet{Plambeck90} model.  A bridge of emission connecting the two
southern clusters led \citet{Greenhill04} to propose that the v=1
masers originate in a nearly edge-on disk rotating about a NE--SW
axis, perpendicular to the previously hypothesized disk, and that the
v=0 SiO emission traces the base of the 18~\kms\ outflow.  This model
did not attempt to explain the high velocity outflow.

Proper motion measurements of source~I \citep{Rodriguez05,Gomez08}
also pose problems for the original model.  These data indicate that
source~I is moving to the SE at $0.007\arcsec$/year, plowing through
the molecular cloud at $\sim 14$~\kms.  In that case a 1000 AU
diameter disk would rapidly be stripped away by the ram pressure of
the ambient gas unless source~I were implausibly massive.

In this paper we report $0.45\arcsec$ (180 AU) resolution images of
the SiO v=0 J=2-1 line obtained with CARMA.  The new maps provide the
best evidence to date that the v=0 SiO emission originates in a NE--SW
outflow from source~I, as in the \citet{Greenhill04} model.  The maps
also suggest that the high velocity outflow is a continuation of this
compact flow.

\section{Observations}

Observations were made with CARMA in the A, B, and C arrays, providing
projected antenna spacings ranging from 5 to 545 k$\lambda$.  The
total integration time on Orion-KL was 3.4 hrs in the A-array (2009
Jan), 3.7 hrs in the B-array (2008 Feb), and 5 hrs in the C-array
(2009 May).

The correlator was configured for simultaneous observations of the
J=2-1 SiO transitions in both the v=0 (86.847 GHz) and v=1 (86.243
GHz) vibrational levels.  For each transition the velocity coverage
was 104~\kms\ and the resolution was 3.4~\kms\ after Hanning
smoothing.  Observations of quasars were used to calibrate phase and
amplitude ripples across the I.F. passband.  After these were removed,
the visibility data were selfcalibrated using a 28~\kms\ wide channel
that included all the bright v=1 maser features.  Only baselines
shorter than 250 k$\lambda$ were used in the selfcalibration because
the maser is slightly resolved on the longest baselines.  The
integration time per data record was 4 sec in the A-array and 10 sec
in the B and C arrays, allowing near-complete removal of atmospheric
phase fluctuations; phase residuals on the maser were typically $\pm
1^{\circ}$ on all baselines after selfcalibration.

Maps of the v=0 SiO line were made with the Miriad data reduction
package.  Uniformly weighting the visibility data yielded an
$0.54\arcsec \times 0.40\arcsec$ synthesized beam at PA $52^{\circ}$.
Because the data were selfcalibrated using the v=1 SiO masers, the
maps are centered on the intensity-weighted mean maser position, which
does not coincide precisely with radio source~I.  An 86 GHz continuum
map made with A-array data from 2009 Jan showed that source~I was
centered $0.01 \pm 0.01\arcsec$ W and $0.03 \pm 0.01\arcsec$ S of the
mean v=1 maser position.  We adjusted the central coordinate of the
maps to place source~I at $05^h35^m14\fs515$,
$-5^{\circ}22\arcmin30\farcs57$, as derived from VLA proper motion
data \citep{Gomez08}.

\section{Results}

\begin{figure} [thb]
\begin{center}
\epsscale{1.2}  
\plotone{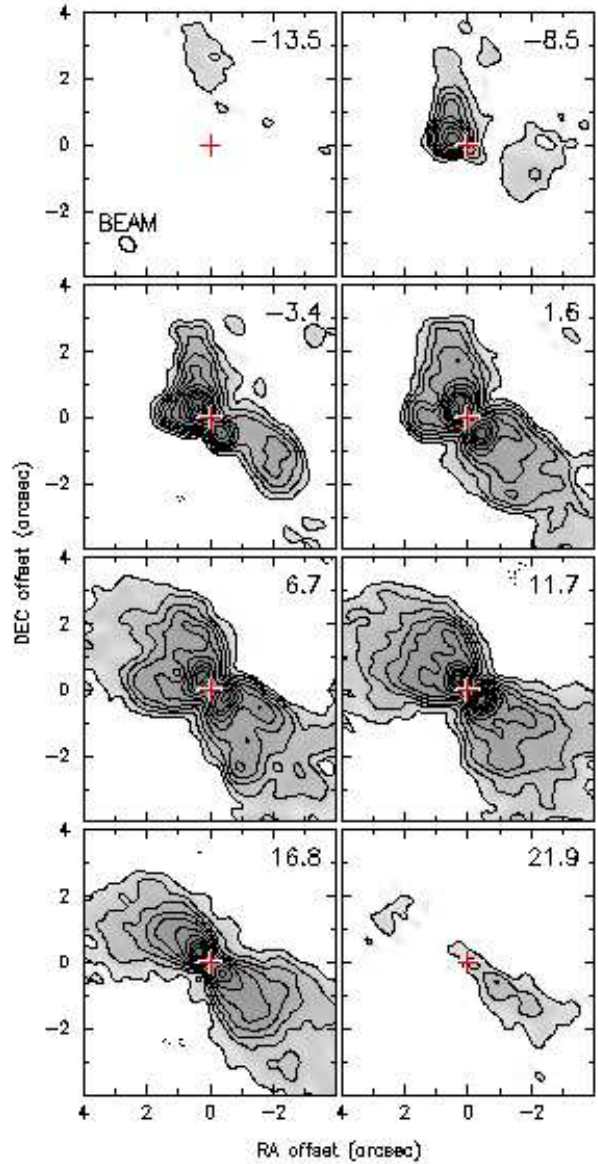}
\caption{
  \label{fig:fig1} Channel maps, 5.06~\kms\ wide, of the SiO v=0 J=2-1
  line toward Orion-KL.  Each box is $8\arcsec$ on a side and is
  labeled with the center LSR velocity. A cross marks the location of
  radio source~I.  The contour levels are $\pm$ 75, 150, 250, 400,
  600~K, then 900 to 3400~K in 500~K steps; 745 K = 1 Jy/beam.  The
  noise level is 12~K, increasing to 19~K in the center channels due
  to limited dynamic range.}
\end{center}
\end{figure}

Figure~\ref{fig:fig1} presents a series of channel maps showing the
v=0 SiO emission in an $8\arcsec \times 8\arcsec$ box centered on
source~I.  The lowest contour on these images is 75 K and the peak is
3800 K.

Although the maps in Figure~\ref{fig:fig1} are consistent with those
published by \citet{Wright95} and \citet{Chandler95}, the higher
quality CARMA images clearly indicate that v=0 SiO emission originates
in a NE--SW bipolar outflow from source~I, as in the
\citet{Greenhill04} model.  The central channels show the
limb-brightened edges of two cones centered on source~I; emission in
the line wings is brightest interior to these cones.  Although the
outflow appears to lie nearly in the plane of the sky, there is a
measurable red-blue asymmetry: redshifted gas at \VLSR = 22~\kms\
appears as a narrow jet to the SW, while blueshifted gas at \VLSR =
$-9$~\kms\ is offset to the NE.  This asymmetry is inconsistent with
the old model in which the emission originates in the turbulent
boundary layer of a disk.

It is clear that the SiO outflow corresponds to the 18~\kms\ H$_2$O
maser outflow identified by \citet{Genzel81}: toward source~I the SiO
line has a fullwidth at zero intensity of $\sim 36$~\kms, and both SiO
and H$_2$O masers extend along a NE--SW axis.  It is likely, however,
that this material was accelerated by a much faster wind.  Source I is
traveling to the SE at 14~\kms\ in the plane of the sky
\citep{Gomez08}, so gas moving outward at only 18~\kms\ would appear
to be swept back into an arc with arms trailing to the NW.

\begin{figure} [thb]
\begin{center}
\epsscale{1.0} 
\plotone{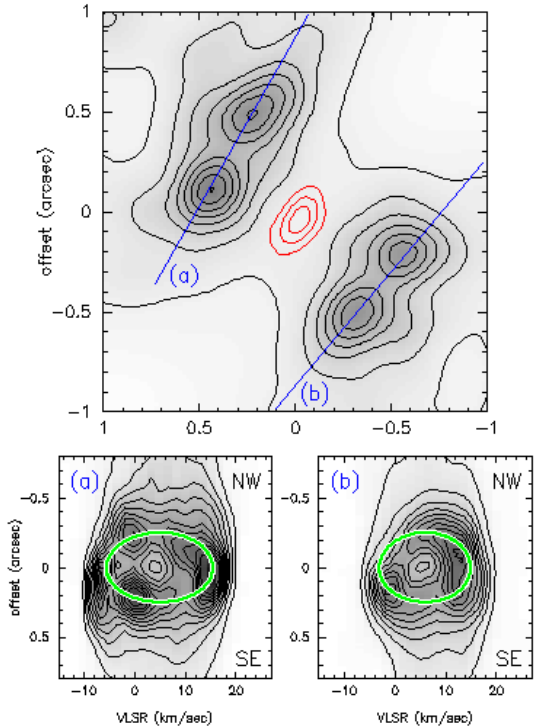}
\caption{
  \label{fig:fig2} ({\it top}) SiO v=0 map in a 10~\kms\ wide channel
  centered at \VLSR = 5~\kms, deconvolved to $0.25\arcsec$ resolution
  as described in the text.  The contour interval is 780~K
  (0.3~Jy/beam), the rms noise is 80~K, and the peak intensity is
  6250~K.  Red contours show the 229 GHz radio continuum from
  source~I, measured at CARMA with an $0.15\arcsec$ synthesized beam
  (Plambeck et al., in preparation).  ({\it bottom}) Position-velocity
  cuts along lines (a) and (b) generated from 3.4 \kms\ resolution
  channel maps, also deconvolved to $0.25\arcsec$ resolution.  The
  contour spacing is 780~K; the rms noise of the channel maps ranges
  from 50--140~K.  Green ellipses show velocities expected for a
  $0.5\arcsec$ diameter ring expanding radially at 10.5~\kms\ in panel
  (a), 9~\kms\ in panel (b).}
\end{center}
\end{figure}

The high signal to noise level in the maps, exceeding 100:1 on the
bright maser spots, makes it feasible to enhance the resolution of the
innermost region of the outflow.  One can fit the position of an
isolated maser feature to an accuracy of $\sim 0.5\,\theta_{\rm
  FWHM}/{\rm SNR}$ \citep{Reid88}, where $\theta_{\rm FWHM}$ is the
apparent FWHM of the source and SNR is the signal-to-noise ratio.
Here the maser spots overlap heavily, however, so fitting their
positions individually is difficult.  Instead, we CLEANed the maps in
the usual way, then convolved the CLEAN components with a
$0.25\arcsec$ FWHM Gaussian restoring beam, about half the width of
the synthesized beam.  Figure~\ref{fig:fig2} shows the resulting SiO
v=0 image at \VLSR = 5~\kms\ in the $2\arcsec \times 2\arcsec$ central
region.  Red contours show the 229 GHz continuum emission from
source~I, which is centered in the waist of the SiO hourglass; SiO
masers in the v=1 vibrational level are clustered along the edges of
the continuum source.

The brightest v=0 SiO masers lie along two bars offset $\sim
0.5\arcsec$ NE and SW of the continuum source.  The ends of the bars
are limb-brightened, suggesting that the masers originate in two
annuli.  Position-velocity cuts through the bars, shown in panels (a)
and (b), hint that these annuli are expanding radially at $\sim
10$~\kms, as modeled by the green ellipses.

\begin{figure} [thb]
\epsscale{1.1} 
\begin{center}
\plotone{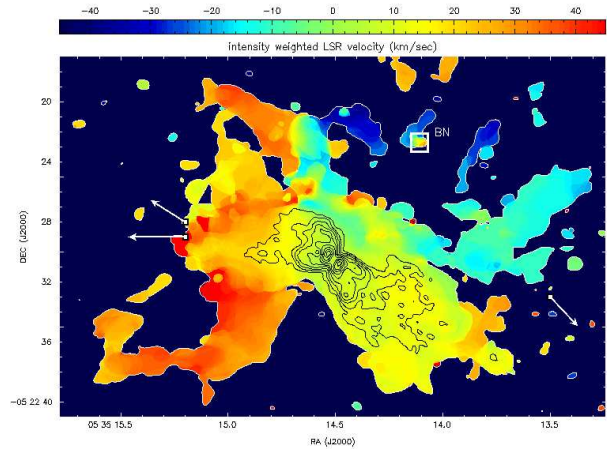}
\caption{
  \label{fig:fig3} Velocity moment map of the v=0 J=2-1 SiO line
  derived from $1\arcsec$ resolution channel maps covering the velocity
  range $-30<$\VLSR$<47$~\kms.  Pixels were averaged into the moment
  map only if they exceeded $3 \times$ the rms noise, which
  ranged from 4--8~K across the channels.  Black contours show the \VLSR = 6.7~\kms\
  channel map from Figure 1, with the same contour intervals.  A white
  box marks the location of the Becklin-Neugebauer Object; radio
  continuum from BN corrupts the velocity moment map in this
  direction.  White arrows show proper motions of the HH objects
  152-228, 152-229, and 135-233 over 150 years \citep{Doi02}.}
\end{center}
\end{figure}

What about the high velocity outflow?  In order to obtain higher
sensitivity for extended emission, we tapered the weighting of the
visibility data to produce a set of channel maps with a $1.13\arcsec
\times 0.93\arcsec$ synthesized beam.  The rms noise is 4~K,
increasing to as much as 8~K in channels with strong maser
emission due to limited dynamic range.  Figure~\ref{fig:fig3} shows
the full extent of the SiO emission detected in these maps, color
coded to indicate the intensity-weighted LSR velocity.  Brightness
temperatures outside the central $8\arcsec \times 5\arcsec$ hourglass
are $<100$~K, hence this emission is likely to be thermal.  As in
previous maps of the high velocity outflow \citep{Chernin96}, the
strongest redshifted emission is E of source~I, while blueshifted gas
is offset to the NW.

Figure~\ref{fig:fig3} suggests that the extended high velocity outflow
is simply a continuation of the compact outflow.  For example, $\sim
3\arcsec$ NE of source~I the compact outflow twists to the E and
points directly toward some of the most redshifted gas in the extended
flow.  The opening angle appears to broaden abruptly at this point,
perhaps as the outflow breaks out of dense gas surrounding source~I.
Two HH objects, 152-228 and 152-229, lie at the tip of the flow;
proper motion measurements indicate that they are moving away from
source~I at 35-50~\kms\ \citep{Doi02}.

\begin{figure}[thb]
\begin{center}
\epsscale{1.1} 
\plotone{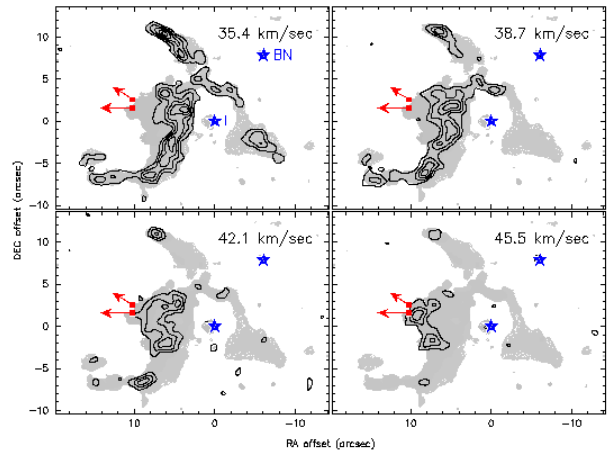}
\caption{
  \label{fig:fig4} SiO v=0 channel maps showing a velocity gradient in
  the redshifted lobe of the high velocity outflow.  The contour
  interval is 12~K; the rms noise is 4~K.  The \VLSR\ of each channel
  is indicated. Gray shading indicates pixels for which there is
  $3\sigma$ emission in at least one channel in the velocity range
  $29<$\VLSR$<52$~\kms.  Blue stars mark the positions of source~I and
  BN.  Red arrows show proper motions of the HH objects 152-228 and
  152-229.}
\end{center}
\end{figure}

Figure~\ref{fig:fig3} suggests that SiO velocities increase away from
source~I.  This is particularly evident in the redshifted gas, as
shown in Figure~\ref{fig:fig4}; note the absence of high velocity gas
within $4\arcsec$ of the star.  There are several possible
explanations for these velocity gradients.

First, SiO emission may originate in dense clumps that are entrained
in a fast stellar wind.  The wind continues to accelerate the clumps
as they move outward.  Evidence for such acceleration is relatively
common in outflows \citep{Stahler94}.

Second, it's possible that the SiO emitting gas was ejected by an
explosive event that flung out material with a range of speeds.  The
fastest-moving ejecta now are farthest from the center.  The system of
shock-excited H$_2$ fingers in Orion appears to have been created by
such an explosive event \citep{Allen93,Doi02}, perhaps as a swept-up
shell was fragmented by a faster-moving wind \citep{Stone95}.

Third, it's possible that the extended gas provides a fossil record of
a precessing outflow.  Currently the outflow axis is NE--SW, with the
(redshifted) SW lobe tipped slightly into the plane of the sky.
Approximately $10\arcsec$ from the star, the outflow axis appears to
be E--W, with the E lobe tipped into the plane of the sky.  If the
true outflow velocity is 100~\kms, this $> 45^{\circ}$ change in
direction took place in 200 years.  Precession occurs because of tidal
interactions in noncoplanar binary systems; periods of a few thousand
years are typical of systems with binary separations of tens of AU
\citep{Terquem99}.  A secondary in an eccentric orbit can change both
the outflow direction and the mass loss rate periodically, as in the
model of CepA presented by \citet{Cunningham09}.

\citet{Zapata09} report that much of the emission from the high
velocity CO line wings in Orion originates from filamentary structures
similar to the H$_2$ fingers, with linear velocity gradients along
them.  \citet{Zapata09} argue that the high velocity flow is not a
classical stellar outflow, but instead was produced by the
disintegration, about 500 years ago, of a multiple stellar system that
contained source~I and BN.  While this scenario is consistent with the
velocity gradients evident in Figure~\ref{fig:fig4} and with the
vaguely filamentary character of the SiO emission away from source~I,
it implies that the apparently continuous transition from the 18 \kms\
flow to the high velocity outflow seen in Figure~\ref{fig:fig3} is
illusory.  We are reluctant to accept that conclusion, and suggest
that perhaps the disintegration of the multiple system triggered an
abrupt increase in the velocity or mass loss rate of the outflow from
source~I, creating the finger system as in the model of
\citet{Stone95}.

\section{Conclusions}

New maps of the SiO line in the v=0 J=2-1 transition toward Orion-KL
provide some of the clearest evidence to date that radio source~I,
thought to be a massive young star, drives a bipolar outflow into the
surrounding molecular cloud along a NE--SW axis.  This outflow
contains multiple water masers and conventionally is referred to as
the 18 \kms\ flow.  The outflow velocity must be substantially greater
than 18~\kms, however, or the outflow lobes would trail back to the NW
owing to the proper motion of source~I.  Close to source~I the SiO v=0
emission is masing.  The strongest masers are clustered in two annuli
offset $\sim 200$ AU from source~I along the central axis of the
outflow.

More extended, higher velocity gas extends along a NW-SE axis, almost
perpendicular to the 18~\kms\ flow.  The SiO maps suggest that this
weakly bipolar outflow is an extension of the 18~\kms\ flow, which may
change direction on time scales of a few hundred years.

\acknowledgements

Support for CARMA construction was derived from the states of
California, Illinois, and Maryland, the Gordon and Betty Moore
Foundation, the Eileen and Kenneth Norris Foundation, the Caltech
Associates, and the National Science Foundation. Ongoing CARMA
development and operations are supported by the National Science
Foundation under a cooperative agreement, and by the CARMA partner
universities.

{\it Facilities:} \facility{CARMA}.

\clearpage


\begin{thebibliography}{25}
\expandafter\ifx\csname natexlab\endcsname\relax\def\natexlab#1{#1}\fi

\bibitem[{{Allen} \& {Burton}(1993)}]{Allen93}
{Allen}, D.~A., \& {Burton}, M.~G. 1993, \nat, 363, 54

\bibitem[{{Beckwith} {et~al.}(1978){Beckwith}, {Persson}, {Neugebauer}, \&
  {Becklin}}]{Beckwith78}
{Beckwith}, S., {Persson}, S.~E., {Neugebauer}, G., \& {Becklin}, E.~E. 1978,
  \apj, 223, 464

\bibitem[{{Chandler} \& {de Pree}(1995)}]{Chandler95}
{Chandler}, C.~J., \& {de Pree}, C.~G. 1995, \apjl, 455, L67+

\bibitem[{{Chernin} \& {Wright}(1996)}]{Chernin96}
{Chernin}, L.~M., \& {Wright}, M.~C.~H. 1996, \apj, 467, 676

\bibitem[{{Cunningham} {et~al.}(2009){Cunningham}, {Moeckel}, \&
  {Bally}}]{Cunningham09}
{Cunningham}, N.~J., {Moeckel}, N., \& {Bally}, J. 2009, \apj, 692, 943

\bibitem[{{Doeleman} {et~al.}(1999){Doeleman}, {Lonsdale}, \&
  {Pelkey}}]{Doeleman99}
{Doeleman}, S.~S., {Lonsdale}, C.~J., \& {Pelkey}, S. 1999, \apjl, 510, L55

\bibitem[{{Doi} {et~al.}(2002){Doi}, {O'Dell}, \& {Hartigan}}]{Doi02}
{Doi}, T., {O'Dell}, C.~R., \& {Hartigan}, P. 2002, \aj, 124, 445

\bibitem[{{Erickson} {et~al.}(1982){Erickson}, {Goldsmith}, {Snell}, {Berson},
  {Huguenin}, {Ulich}, \& {Lada}}]{Erickson82}
{Erickson}, N.~R., {Goldsmith}, P.~F., {Snell}, R.~L., {Berson}, R.~L.,
  {Huguenin}, G.~R., {Ulich}, B.~L., \& {Lada}, C.~J. 1982, \apjl, 261, L103

\bibitem[{{Genzel} {et~al.}(1981){Genzel}, {Reid}, {Moran}, \&
  {Downes}}]{Genzel81}
{Genzel}, R., {Reid}, M.~J., {Moran}, J.~M., \& {Downes}, D. 1981, \apj, 244,
  884

\bibitem[{{Gezari} {et~al.}(1998){Gezari}, {Backman}, \& {Werner}}]{Gezari98}
{Gezari}, D.~Y., {Backman}, D.~E., \& {Werner}, M.~W. 1998, \apj, 509, 283

\bibitem[{{G{\'o}mez} {et~al.}(2008){G{\'o}mez}, {Rodr{\'{\i}}guez}, {Loinard},
  {Lizano}, {Allen}, {Poveda}, \& {Menten}}]{Gomez08}
{G{\'o}mez}, L., {Rodr{\'{\i}}guez}, L.~F., {Loinard}, L., {Lizano}, S.,
  {Allen}, C., {Poveda}, A., \& {Menten}, K.~M. 2008, \apj, 685, 333

\bibitem[{{Greenhill} {et~al.}(1998){Greenhill}, {Gwinn}, {Schwartz}, {Moran},
  \& {Diamond}}]{Greenhill98}
{Greenhill}, L.~J., {Gwinn}, C.~R., {Schwartz}, C., {Moran}, J.~M., \&
  {Diamond}, P.~J. 1998, \nat, 396, 650

\bibitem[{{Greenhill} {et~al.}(2004){Greenhill}, {Reid}, {Chandler}, {Diamond},
  \& {Elitzur}}]{Greenhill04}
{Greenhill}, L.~J., {Reid}, M.~J., {Chandler}, C.~J., {Diamond}, P.~J., \&
  {Elitzur}, M. 2004, in IAU Symposium, Vol. 221, Star Formation at High
  Angular Resolution, ed. M.~G. {Burton}, R.~{Jayawardhana}, \& T.~L. {Bourke},
  155--+

\bibitem[{{Kwan} \& {Scoville}(1976)}]{Kwan76}
{Kwan}, J., \& {Scoville}, N. 1976, \apjl, 210, L39

\bibitem[{{Menten} {et~al.}(2007){Menten}, {Reid}, {Forbrich}, \&
  {Brunthaler}}]{Menten07}
{Menten}, K.~M., {Reid}, M.~J., {Forbrich}, J., \& {Brunthaler}, A. 2007, \aap,
  474, 515

\bibitem[{{Plambeck} {et~al.}(1990){Plambeck}, {Wright}, \&
  {Carlstrom}}]{Plambeck90}
{Plambeck}, R.~L., {Wright}, M.~C.~H., \& {Carlstrom}, J.~E. 1990, \apjl, 348,
  L65

\bibitem[{{Reid} {et~al.}(1988){Reid}, {Schneps}, {Moran}, {Gwinn}, {Genzel},
  {Downes}, \& {Roennaeng}}]{Reid88}
{Reid}, M.~J., {Schneps}, M.~H., {Moran}, J.~M., {Gwinn}, C.~R., {Genzel}, R.,
  {Downes}, D., \& {Roennaeng}, B. 1988, \apj, 330, 809

\bibitem[{{Rodr{\'{\i}}guez} {et~al.}(2005){Rodr{\'{\i}}guez}, {Poveda},
  {Lizano}, \& {Allen}}]{Rodriguez05}
{Rodr{\'{\i}}guez}, L.~F., {Poveda}, A., {Lizano}, S., \& {Allen}, C. 2005,
  \apjl, 627, L65

\bibitem[{{Sandstrom} {et~al.}(2007){Sandstrom}, {Peek}, {Bower}, {Bolatto}, \&
  {Plambeck}}]{Sandstrom07}
{Sandstrom}, K.~M., {Peek}, J.~E.~G., {Bower}, G.~C., {Bolatto}, A.~D., \&
  {Plambeck}, R.~L. 2007, \apj, 667, 1161

\bibitem[{{Stahler}(1994)}]{Stahler94}
{Stahler}, S.~W. 1994, \apj, 422, 616

\bibitem[{{Stone} {et~al.}(1995){Stone}, {Xu}, \& {Mundy}}]{Stone95}
{Stone}, J.~M., {Xu}, J., \& {Mundy}, L.~G. 1995, \nat, 377, 315

\bibitem[{{Terquem} {et~al.}(1999){Terquem}, {Eisl{\"o}ffel}, {Papaloizou}, \&
  {Nelson}}]{Terquem99}
{Terquem}, C., {Eisl{\"o}ffel}, J., {Papaloizou}, J.~C.~B., \& {Nelson}, R.~P.
  1999, \apjl, 512, L131

\bibitem[{{Wright} {et~al.}(1995){Wright}, {Plambeck}, {Mundy}, \&
  {Looney}}]{Wright95}
{Wright}, M.~C.~H., {Plambeck}, R.~L., {Mundy}, L.~G., \& {Looney}, L.~W. 1995,
  \apjl, 455, L185+

\bibitem[{{Zapata} {et~al.}(2009){Zapata}, {Schmid-Burgk}, {Ho}, {Rodriguez},
  \& {Menten}}]{Zapata09}
{Zapata}, L.~A., {Schmid-Burgk}, J., {Ho}, P.~T.~P., {Rodriguez}, L.~F., \&
  {Menten}, K. 2009, arXiv:0907.3945v1

\bibitem[{{Zuckerman} {et~al.}(1976){Zuckerman}, {Kuiper}, \& {Rodriguez
  Kuiper}}]{Zuckerman76}
{Zuckerman}, B., {Kuiper}, T.~B.~H., \& {Rodriguez Kuiper}, E.~N. 1976, \apjl,
  209, L137

\end{thebibliography}
\end{document}